\documentstyle[12pt,aas2pp4]{article}

\slugcomment{Accepted for publication in the Astrophysical 
Journal Letters, November 1997.}

\lefthead{Phillipps et al.}

\righthead{Dwarf Spheroidal Galaxies in the Virgo Cluster} 

\begin{document}

\title{Dwarf Spheroidal Galaxies in the Virgo Cluster}

\author{S. Phillipps}
\affil{Astrophysics Group,  Department of Physics, University of Bristol, \\
    Tyndall Avenue, Bristol, BS8 1TL, U.K.}

\author{Q. A. Parker}
\affil{Anglo-Australian Observatory, Siding Spring, Coonabarabran, 
    New South Wales, Australia} 

\and

\author{J. M. Schwartzenberg and J. B. Jones}
\affil{Astrophysics Group,  Department of Physics, University of Bristol, \\
    Tyndall Avenue, Bristol, BS8 1TL, U.K.}

\author{Accepted for publication in the {\it Astrophysical 
Journal Letters}.}

\begin{abstract}
We present a study of the smallest and faintest galaxies found in a very deep
photographic R band
survey of regions of the Virgo Cluster, totalling over 3 square degrees,
made with the UK Schmidt Telescope. 
The objects we detect have the same physical sizes and surface brightnesses as 
Local Group dwarf spheroidal galaxies.
The luminosity function of these extremely low luminosity galaxies (down
to $M_{R} \simeq -11$ or about $5 \times 10^{-5} L_{\star}$) 
is very steep, with a power law slope $\alpha \simeq -2$,
as would be expected in many theories of galaxy formation via hierarchical
clustering, supporting previous observational
evidence at somewhat higher luminosities in other clusters.
\end{abstract}

\keywords{galaxies: clusters: individual (Virgo) --- 
   galaxies: elliptical and lenticular, cD ---
   galaxies: evolution  ---
   galaxies: luminosity function, mass function ---
   galaxies: photometry}

\twocolumn

\section{Introduction}
Recent observations of rich clusters have indicated that the galaxy luminosity
function (LF) may turn up at the faint end (e.g., Driver et al. 1994; Mohr et al. 1996;
Smith et al. 1997; Wilson et al. 1997; Trentham 1997). From a Schechter (1976)
function slope of $\alpha \simeq -1$, there appears to be a steepening to
$\alpha \simeq -1.5$ to --1.8 below about $0.04 L_{\star}$
(roughly $M_{R} \simeq -18$ or $M_{B} \simeq -16.5$ for $H_{0} = 75$
kms$^{-1}$Mpc$^{-1}$). However the LF is
still largely unknown faintwards of $M_{R} \simeq -15.5$ ($M_{B} \simeq -14$),
so the overall 
contribution of dwarf galaxies to the cluster population remains uncertain.
The Local Group provides the only well studied sample of such faint
galaxies and it appears that below about $M_{R} = -15.5$ ($0.004 L_{\star}$)
the galaxies are
virtually all dwarf spheroidals (van den Bergh 1992a).
These objects span the magnitude range from about $M_{R} \simeq -14.5$ (Fornax)
to $\simeq -8.5$ (Carina), and indicate a rather flat
luminosity distribution (van den Bergh 1992b). 

In this paper we present a deep photographic survey of significant
areas within the Virgo Cluster which span a range of (giant) galaxy densities.
The survey, based on digitally stacking UK Schmidt Telescope films (cf.
Bland-Hawthorn, Shopbell \& Malin 1993),
extends the earlier, seminal, survey of Binggeli, Sandage \&
Tammann (1985), reaching roughly 3 magnitudes beyond their 
completeness limits.
We are able to detect large numbers of faint galaxies, presumably dwarf
spheroidals, at magnitudes down to $M_{R} \simeq -11$ for an
assumed Virgo distance of 18 Mpc (e.g., Jacoby et al. 1992; for consistency 
we also adopt their value of $H_{0} = 75$ km s$^{-1}$ Mpc$^{-1}$ where
required).

\section{The Data and Image Detection}

The data used here are part of a larger photographic survey of the Virgo 
Cluster (Schw\-art\-zenberg, Phillipps \& Parker 1995a) using the extremely
fine grained, highly
efficient Tech Pan films on the 1.2m UK Schmidt Telescope 
(Phillipps \& Parker 1993).
Six individual long (1 to 1.5 hour) exposures of the same area, the
South East quadrant
of the Virgo Cluster, were scanned with the SuperCOSMOS automatic measuring
machine at the Royal Observatory Edinburgh (Miller et al. 1992). 
For convenience, nine
separate scan regions 6840 pixels square were created from each film. The
pixel scale is 10 microns or $0.67''$ giving a total area for each scan
$\simeq 1.3^{\circ} \times 1.3^{\circ}$. Two of these scan regions
are considered in the present paper, one
near the cluster centre, one further out. Note that, while large in itself,
the $\simeq 3.2$ square degrees covered here is only about 1/40 of the
full cluster survey area.

The scans from the six separate films
were sky subtracted by using a $256 \times 256$ pixel spatial
median filtered version of the data themselves, then matched in intensity
by comparing images of a number of calibrating galaxies and 
median stacked (see Schwartzenberg, Phillipps \& Parker 1996 for details).
Median stacking has equivalent noise reduction to simple co-addition
and is highly effective in removing artefacts (e.g., due to satellite trails, 
dust particles adhering to the emulsion and so forth) which affect only one film in the stack. Absolute
calibration was via comparison of the images of some brighter
galaxies with published CCD photometry (from Gallagher \& Hunter 1989), as
described by Phillipps \& Parker (1993).
The final stacked data have an equivalent exposure time of about 7 hours,
and the high efficiency of the films, approaching 10\% (Parker et al. 1997),
results in a pixel-to-pixel sky noise $\sigma_{sky}$
equivalent to 26.2 $R$ magnitudes
per square arc second (henceforth $R\mu$).

Galaxy (and star) images were automatically recovered from the stacked data via
a connected pixel algorithm ({\sc PISA}, Draper 1993), 
using a detection threshold
$2\sigma_{sky}$ above the sky background,
or $25.45 R\mu$, and a minimum area of 25 pixels (11
square arc seconds). Each image thus has a minimum S/N of 10 and has a
magnitude $R \leq 22$. (In principle, $3 \sigma$ detection of images of
size around $2''$ would reach a magnitude limit of about $R = 24$). Around
28,000 images are detected in each field and the large minimum area
ensures that few are spurious. This was confirmed by comparison of a
small area of the photographic data with a CCD image taken on the
Anglo-Australian Telescope; 160 of 162 images visible on the film
were matched on the CCD frame, including {\it all} those in the `refined' samples used below. 

\section{Results}

The two areas chosen for this study are centred close to M87 and $3.^{\circ}1$
to its south south east, enabling both the cluster core and a more typical
cluster region to be investigated. 
The area immediately around M87 was not searched
due to its effectively higher background light level, so the inner 
and outer surveyed
fields cover 1.58 and 1.61 square degrees, respectively.
Once the raw catalogues were produced, as above, we `refined' them by requiring
that our images met certain criteria aimed at isolating low surface brightness
cluster dwarfs. In particular we kept only those images whose isophotal
sizes and isophotal magnitudes were consistent with them having
exponential profiles (characteristic of virtually all dwarfs; see Binggeli
\& Cameron 1991) of scale size $a \geq 2''$ and central surface brightness
$\mu_{0} \geq 22R\mu$ (cf. Figure 1 of Schwartzenberg et al. 1995b). This
reduced the number of potential Virgo Cluster medium to low surface brightness 
dwarf galaxy candidates to approximately 17,000 (from 56,000 images of all types). Note that we
do not separately remove stars as these should disappear along with the higher
surface brightness galaxies. The data set includes galaxies
with $\mu_{0}$ down to about 25.2 $R\mu$, 
but is complete (in the sense that even a
$2''$ scale size gives images exceeding our area limit) only to $24.5 R\mu$. Further details of other image parameters are given by
Schwartzenberg (1996), but here we concentrate solely on the magnitudes,
though note that we use `total' magnitudes calculated from the measured
$a$ and $\mu_{0}$.

In principle it is possible for an LSBG sample to contain cosmologically 
dimmed normal surface brightness giants at large redshifts or large non-cluster
LSBGs in the background. The former would generally appear much smaller than 
our detection limit (with $a \leq 1''$, cf. Windhorst et al. 1994), while
the latter are relatively rare (see Schwartzenberg et al. 1995b) and can be
subtracted statistically (Turner et al. 1993). Nevertheless, in order 
to reduce such contamination problems to a minimum, we have again refined
our sample to include only the $\simeq$ 4000 objects with $a \geq 3''$. These 
images will also be less affected by seeing; even if the scale
lengths are slightly increased by the blurring (and the relatively moderate 
resolution), the central surface brightness will be decreased to
compensate, leading to little error in the derived total
magnitudes. In effect we will have merely the LF of
galaxies limited at a marginally smaller physical size than would have been
the case in the absence of seeing. 

Since we have pre-selected our dwarf LSBG sample in terms of scale size and 
central surface brightness, we cannot simply subtract standard number counts 
for the entire population of field galaxies
(e.g., Metcalfe et al. 1995) from the magnitude distribution we
obtain, in order to arrive at the cluster LF. We have therefore made a 
subtraction based on the corresponding distribution of {\em field} LSBGs
parameters found by Schwartzenberg et al. (1995b). This correction turns
out to be quite small compared to our total LSBG numbers (a few percent), so
is not critical to our final LF, for the simple reason that most background
LSBGs appear much smaller than our cluster LSBG candidates (cf.
Karachentsev et al. 1995).

Figures 1a and 1b illustrate the LFs for the outer (895
galaxies) and inner (675 galaxies) cluster regions, respectively. We show
here only those galaxies with $a \geq 3''$ {\it and} $22.0 \leq \mu_{0} \leq 24.5 R\mu$,
the region of parameter space for which we have a complete and minimally
contaminated sample. Note that this is {\it not} a magnitude limited sample.
For instance, we already begin to lose any galaxies with smaller scale 
sizes at $M_{R} \simeq -13$, whereas our faintest objects have $M_{R} 
\simeq -11$.
Of course, even the loss of small objects at faint $M_{R}$ 
may not be the whole
story as far as the LF goes, since there may exist higher surface brightness dwarfs
than we are allowing for (perhaps preferentially at bright $M_{R}$), 
and we will be missing any even lower surface brightness objects
at {\it all} $M_{R}$.
Indeed, recall that we have many candidates for smaller or lower
surface brightness galaxies in our original overall sample (see also
Schwartzenberg 1996). 

The LFs plotted for the two regions 
in Figure 1 are very similar so an overall LF for
the samples can be used. It is clear that the LF
is again steep, as in the papers discussed in Section 1, confirming
earlier suggestions for Virgo itself by Impey, Bothun \& Malin (1988)
and Tyson \& Scalo (1988). A least squares fit to
the combined data 
gives a power law slope for the range $15.5 < R < 20.0$ (roughly --16 to --11.5
in $M_{R}$)
of $\alpha = -2.26 \pm 0.13$. (Fitting to the individual LFs over the same
magnitude range gives  $\alpha = -2.26 \pm 0.14$ and $-2.18 \pm 0.12$).
Note that if we restrict attention to the very faint galaxies,
$R > 18$ ($M_{R} > -13.5$), the 
steepening is even more dramatic, $\alpha \simeq -2.5$. (The turn up appears
clearer in the `outer' field, Figure 1a).
The amplitudes of the LF for the two separate
areas are also similar (in galaxies per magnitude bin per square degree),
with the core sample actually having the lower projected density by a factor $\simeq 0.8$. LSBGs may be adversely affected by the presence of the 
giant galaxy M87 in the cluster centre region; 
Thompson \& Gregory (1993) have previously found a similar effect in the
core of the Coma Cluster. Note, though,
that with the very steep LF slope, a 
relatively small zero point offset in the calibration between fields,
for instance, can have a
significant effect on the numbers. For instance an error of 
$\Delta m$ = 0.$^{m}$1 
generates a difference of a factor $10^{0.4 (\alpha + 1) \Delta m} \simeq$ 1.12.
Such errors would make little difference to the shape of the derived 
LF, since the background contamination is so small.
At the bright end, the number of detected LSBGs
is in good agreement with that expected from the Binggeli, Sandage \&
Tammann (1985, 1988) Virgo Cluster LF, given the small numbers of these
objects in our samples (see Figure 1b). 
It is clear that our LF departs from theirs
at the expected point where incompleteness and lack of very low surface
brightness objects starts to affect their sample, beyond $R \simeq 17$.
(We assume here typical early type galaxy colours, $B-R = 1.5$, for
the dwarf spheroidals; if they
actually have bluer colours, as often seen in low surface
brightness galaxies, this would slightly improve the match).

\section{Conclusions}

By co-adding very deep UKST photographic films we have been able to
probe the dwarf population of the Virgo Cluster down to $M_{R} \simeq -11$
($\simeq 5 \times 10^{-5} L_{\star}$).
The central surface brightness limit for our sample is
$25 R\mu$, corresponding roughly to $26.5 B\mu$ for early type galaxy colours.
In both luminosity and surface brightness
this is thus one of the deepest surveys yet performed. In
particular, our limits allow us to survey
well into the regime of the dwarf spheroidal galaxies (Irwin \&
Hatzidimitriou 1995); the luminosity limit is 25 times fainter
than the Fornax dwarf, for instance. We have therefore been able to gather
by far the largest sample of dwarf ellipticals/dwarf spheroidals currently
known. Of course, in the absence of redshifts, these `detections' are 
on a statistical basis only. However, given the paucity of LSBs of moderate to
large angular size in the general field, we are confident that the large
majority of our candidates are genuine cluster dwarfs.

The luminosity function of the dwarfs is very steep, with $\alpha \simeq -2$,
confirming values found over much more limited magnitude ranges in other
clusters (e.g., Smith et al. 1997; Trentham 1997). 
Bernstein et al. (1995) reached
similarly faint levels to those discussed here
with very deep CCD imaging of a very small area at the core of the
Coma cluster (to $M_{R} = -11.4$). They found a less steep LF than
most other deep surveys, $\alpha \simeq -1.3$, but the
centre of Coma may be a rather special
environment. Though their surveys are less deep, Biviano et al. (1995) 
and Thompson \& Gregory (1993) find steeper slopes
for parts of Coma further from the centre.

A steep slope, $\alpha \simeq -2$, is as
expected generally in any hierarchical structure formation model
(eg. White \& Frenk 1991; Blanchard, Valls-Gabaud \& Mamon 1992;
Evrard, Summers \& Davis 1994; Frenk et al. 1996; Kauffmann Nusser \& 
Steinmetz 1997). Note that since we are observing at long wavelengths
($R$ band), and in any case we expect most of our objects to be dwarf
ellipticals with little or no recent star formation, our LF shape
should closely match that of
the more fundamental (baryonic) mass function, allowing a simpler comparison
with theoretical models. There is a suggestion that the dwarf LSBG
to giant galaxy ratio is smaller in the cluster core than further out. 
Analysis of the whole cluster survey area should allow us to quantify this
in more detail (see also Phillipps et al. 1997).

We might, finally, note  
that much smaller and fainter galaxies can be {\it detected} in our data 
than are present in our photometric samples,
Indeed, we can reach down
to about $M_{R} \simeq -8.5$ at the distance of Virgo, the same as the
Carina dwarf, the lowest luminosity system currently known. 
Unfortunately, though, these
images are indistinguishable from those of the (very numerous) general
background galaxies. However, we should still be able to estimate their
numbers through a comparison with an identically observed non-cluster
field. This work will be reported in a subsequent paper. At some point
we should certainly expect to see a turn over in the LF, since $\alpha =
-2$ is the critical value at which the integrated galaxy light formally
diverges. If the currently found slopes in the range --2 to --2.5
were to continue down to, say, $M_{R} = -8$, then the dwarf galaxies
fainter than $M_{R} = 16$ would contain approximately 0.1 to 1.0 
times as much light as the 
brighter galaxies. For a constant $M/L$ (or perhaps more reasonably, a
fixed {\it baryonic} $M/L$) this would obviously increase the total 
mass in cluster galaxies by a factor between 1.1 and 2.

\acknowledgements
We would like to thank the UK Schmidt Telescope for the provision of the
usual excellent photographic material and the SuperCOSMOS group at ROE for
scanning them.
SP and JBJ thank the Royal Society and the UK PPARC, respectively, for
financial support. 
\\[10mm]

\clearpage
\onecolumn

\begin{figure}
\epsscale{.5}
\plotone{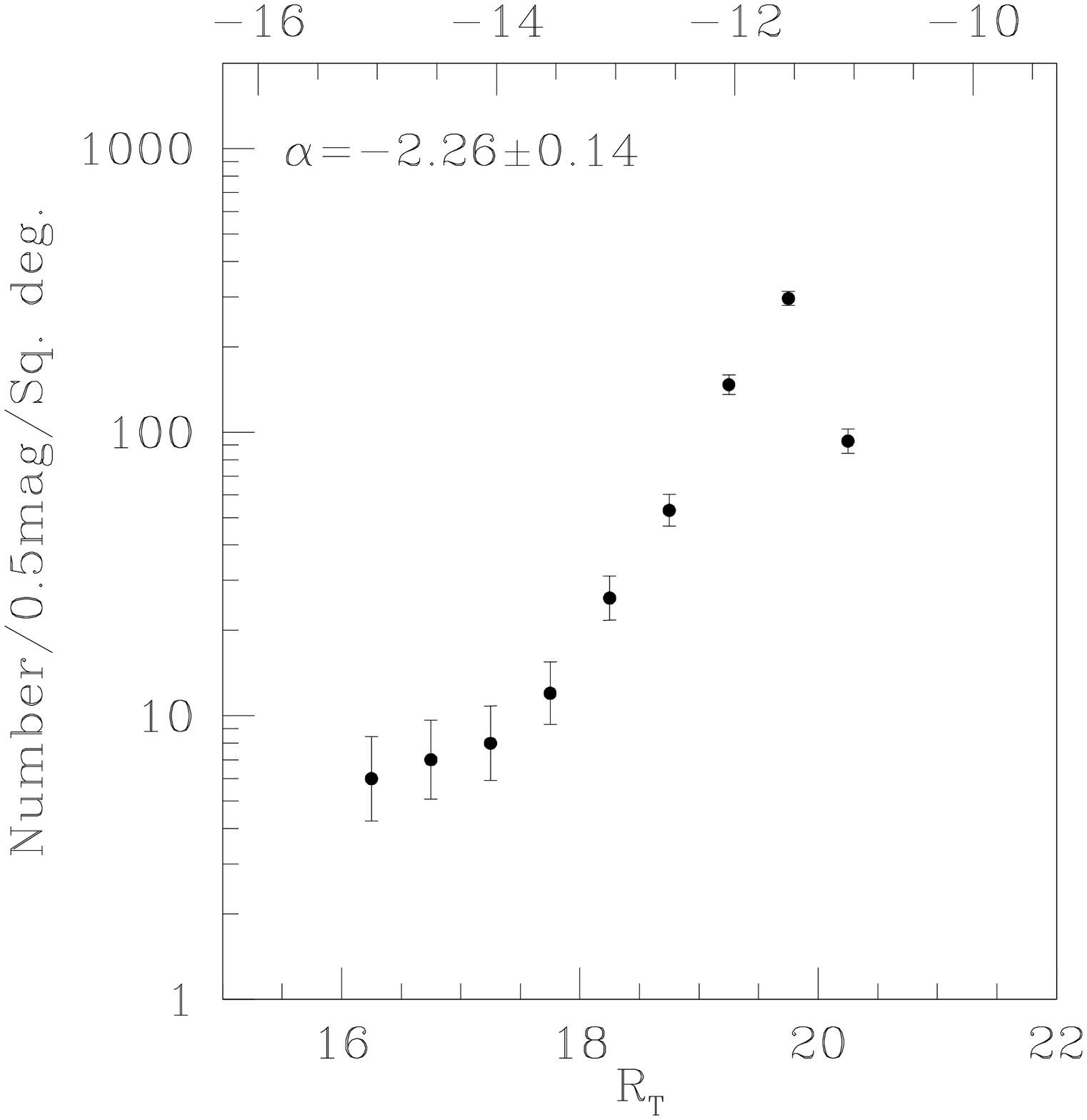}
\plotone{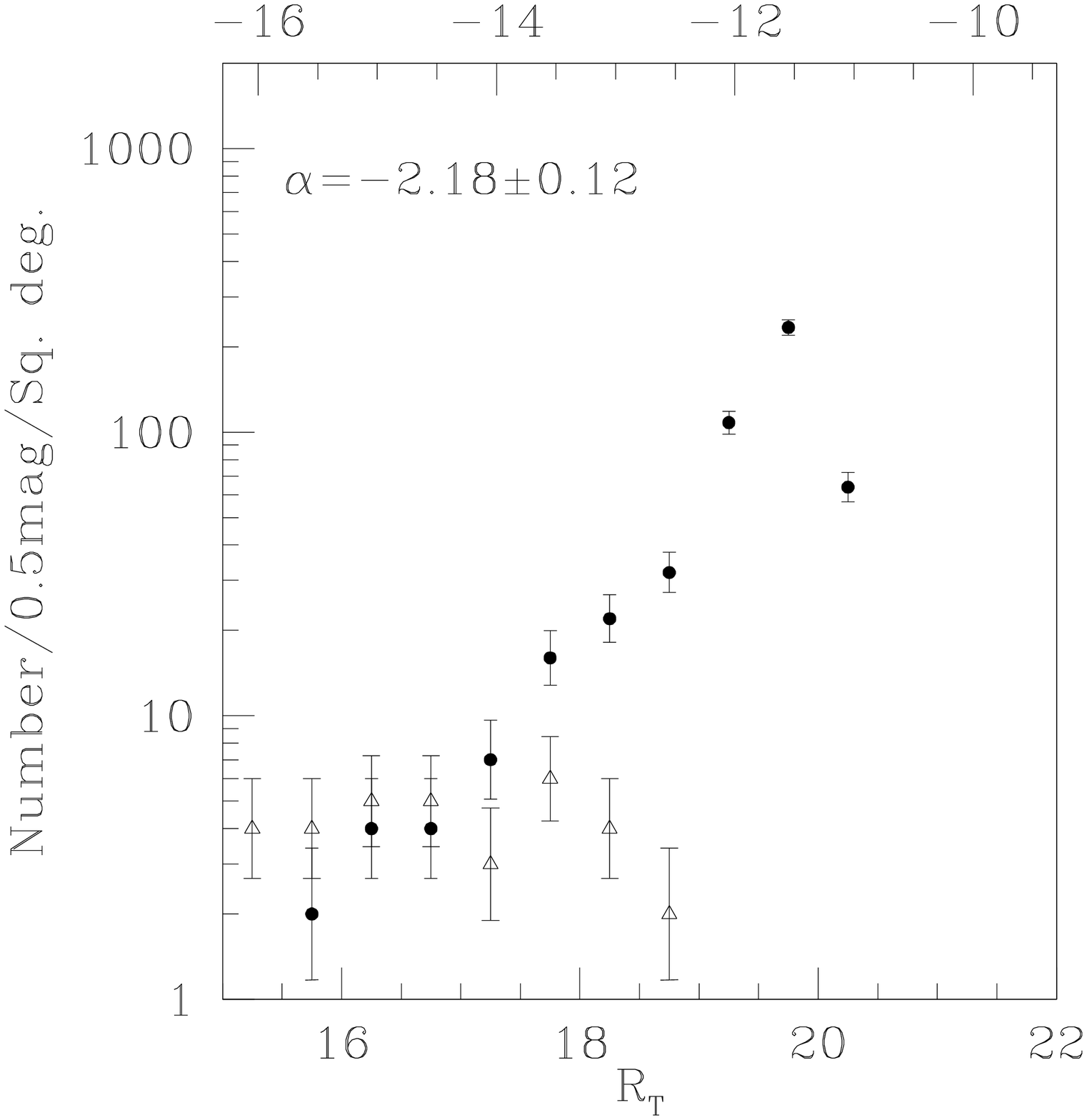}
\caption{The absolute magnitude distribution of 
Virgo LSBGs (as defined in the text), 
for the (a) outer and (b) inner area samples. Error bars 
shown are based on Poissonian statistics. The luminosity function 
of Sandage et al. (1985) is also shown (open triangles) for the 
Virgo Cluster Catalog galaxies which overlap with the inner field.
(We assume for this comparison $B-R = 1.5$.) 
\label{fig1}} 
\end{figure}
 
\end{document}